\documentclass[%
reprint,
 amsmath,amssymb,
 aps,
]{revtex4-1}
\usepackage{amsmath}
\usepackage{}

\usepackage{graphicx}% Include figure files
\usepackage{dcolumn}% Align table columns on decimal point
\usepackage{bm}
\usepackage{epstopdf}
\usepackage{epsfig}% bold math
\usepackage[colorlinks,
            linkcolor=blue,
            anchorcolor=blue,
            citecolor=blue
            ]{hyperref}
\usepackage{color}

\begin{document}

%\preprint{APS/123-QED}

\title{Fundamental Limits for Reciprocal and non-Reciprocal non-Hermitian Quantum Sensing}% Force line breaks with \\
\author{Liying Bao$^{1,2}$, Bo Qi$^{1,2,\star}$,  Daoyi Dong$^{3}$, Franco Nori$^{4,5}$\\
$^{1}$\textit{Key Laboratory of Systems and Control, Academy of Mathematics and Systems Science, Chinese Academy of Sciences, Beijing 100190, P. R. China}\\
$^{2}$\textit{University of Chinese Academy of Sciences, Beijing 100049, P. R. China}\\
$^{3}$\textit{School of Engineering and Information Technology, University of New South Wales, Canberra, ACT 2600, Australia}\\
$^{4}$\textit{Theoretical Quantum Physics Laboratory, RIKEN, Saitama, 351-0198, Japan}\\
$^{5}$\textit{Physics Department, The University of Michigan, Ann Arbor, Michigan 48109, USA}\\
$^\star$\textit{qibo@amss.ac.cn}}

\date{\today}% It is always \today, today,

\begin{abstract}
Non-Hermitian dynamics has been widely studied to enhance the precision of quantum sensing; and non-reciprocity  can be a powerful resource for non-Hermitian quantum sensing, as non-reciprocity allows to arbitrarily exceed the fundamental bound on the measurement rate of any  reciprocal sensors. Here we establish fundamental limits on signal-to-noise ratio for reciprocal and non-reciprocal non-Hermitian quantum sensing. In particular,  for two-mode linear systems with two coherent drives, an approximately attainable uniform bound on the best possible measurement rate per photon is derived for both reciprocal and non-reciprocal sensors. This bound is only related to the coupling coefficients and, in principle, can be made arbitrarily large. Our results thus demonstrate that a conventional reciprocal sensor with two drives can simulate any non-reciprocal sensor.  This work also demonstrates a clear signature on  how the excitation signals affect the signal-to-noise ratio in non-Hermitian quantum sensing.

%\begin{description}
%\end{description}
\end{abstract}

%\pacs{Valid PACS appear here} % PACS, the Physics and Astronomy
\maketitle

%\tableofcontents

\section{Introduction}

An important task in quantum science and technology is to investigate the precision limit of quantum sensing and devise protocols to attain it.
Non-Hermitian dynamics \cite{Ozdemir2019,El-Ganainy2018,Pan2020,Scheibner2020,Ashida2020,Liuyl2017,Huai2019,ZhangJing2018,Chen2017,Leykam2017,Feng2017,
Liu2019,Zhang2019,Lau2018,Chen2019,Langbein2018,Bliokh2019,Moiseyev2011, Bender2007,Mudry1998,Demange2012,Chu2020,Cao2020,Chen2016} has attracted much interest in recent years for their possibility in enhancing quantum sensing. Most of the key results refer to the intriguing non-Hermitian degeneracy property  known as the exceptional point (EP), at which not only the eigenenergies but also the eigenstates coalesce \cite{Berry2004,Jing2014,Hodaei2017,Liertzer2012,Dembowski2001,Zhen2015,Xu2016,Wiersig2014,Rotter2014,Liu2016,Heiss2012,
Heiss2000,Heiss2004,Sunada2017,Wiersig2016,Ren2017,Seyranian2005,Heiss1999}.  Near the EP, the eigenenergies have a diverging susceptibility on small parameter changes, which is leveraged for sensing  weak signals. When utilizing  EP sensors, fine tunings of the system parameters are needed. Moreover, the real effect of EP sensors should be assessed by taking full account of  noise effects and/or  realistic measurements owing to the fact that the coalescence of eigenstates may suppress the diverging susceptibility of eigenenergies \cite{Zhang2019,Lau2018,Chen2019}.
%This means that if the coalescence of $n$ levels occurs,  a small parameter change $\varepsilon$ is able to lead to an eigenenergy splitting $\varepsilon^{\frac{1}{n}}$, and the susceptibility on the parameter change diverges at $\varepsilon=0$ since $$\frac{d}{d\varepsilon}\varepsilon^{\frac{1}{n}}\propto\varepsilon^{\frac{1-n}{n}}.$$

Recently there have been several theoretical results calculating the signal-to-noise ratio (SNR) and the quantum Fisher information of EP sensors \cite{Zhang2019,Lau2018,Chen2019, Langbein2018}. It has been demonstrated in \cite{Lau2018} that amplification or gain is a necessary ingredient for enhancing signal powers but there is no fundamental utility using an EP sensor. Furthermore,  non-reciprocity \cite{Lau2018,Sounas2017,Zhangjing2015,Metelmann2015,Fang2017,Bernier2017,Ozdemir2014}, where the magnitude of the coupling between two modes has directionality, was demonstrated to be a powerful resource for quantum sensing. This was concluded by first deriving fundamental bounds on the measurement rate that constrains any reciprocal two-mode systems including reciprocal EP sensors, and then demonstrating that breaking reciprocity allows to arbitrarily exceed the bounds restricting reciprocal sensors \cite{Lau2018}. It is worth stressing that non-reciprocity has nothing to do with EP.

Inspired by \cite{Lau2018}, we derive fundamental bounds on the SNR for linear coupled-mode non-Hermitian systems with two coherent drives,  instead of the one drive used in previous works.  Focusing on two-mode systems, we show that with two coherent drives, a uniform bound on the best possible measurement rate per photon, which determines the best possible rate of SNR growing in time per photon, can be derived for both reciprocal and non-reciprocal sensors. This bound is approximately attainable and only related to the  coupling coefficients and can, in principle, be made arbitrarily large. Our results show that \textit{conventional reciprocal sensors with two coherent drives}, which can be relatively easy to implement with current technology, \textit{can simulate any non-reciprocal sensor}. Moreover, the introduction of two drives provides a clear signature on understanding  how the SNR relates to the excitation signals.

The paper is organized as follows. In Section II, we describe the  generic non-Hermitian setup in terms of the Heisenberg-Langevin equations  and depict the SNR and the measurement rate per photon to be used explicitly. A general bound on non-Hermitian sensing is derived in Section III, and then we apply the result to two-mode systems in Section IV. We then investigate how the drive frequencies affect the measurement rate per photon in Section V.  Section VI concludes the paper.

\begin{figure}[!htb]
  \centering
  \includegraphics[width=\hsize]{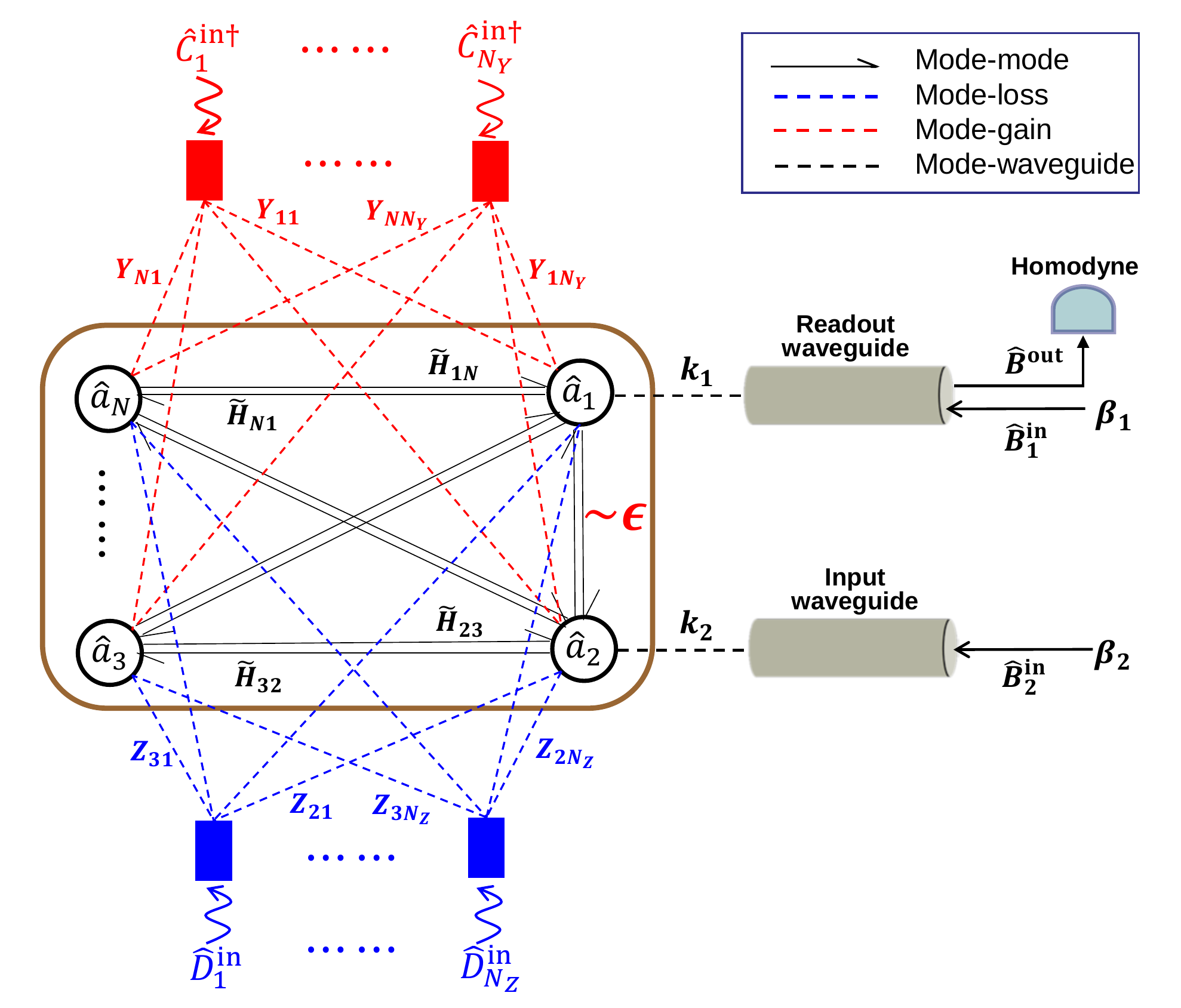}
  \caption{(Color online) A generic non-Hermitian linear mode setup. The circles denote resonant linear modes that interact according to the effective non-Hermitian Hamiltonian $\tilde{H}[\epsilon]$, where the parameter $\epsilon$ is an infinitesimal disturbance to be measured. The linear modes are coupled to dissipative baths via gain (red) and loss (blue) processes.   Two coherent drives are injected into mode $1$ ($\hat{a}_1$) and mode $2$ ($\hat{a}_2$) through  two waveguides, respectively. The reflected field from mode $1$ is measured by a homodyne detection. }
\end{figure}

\section{Non-Hermitian sensing }
\subsection{A generic non-Hermitian setup}
A generic linear non-Hermitian sensing setup is shown in Fig.~1. This is a generalized version of non-Hermitian sensing systems in previous works \cite{Ozdemir2019,Hodaei2017,Chen2017,Lau2018,Wiersig2014,Sunada2017}, which takes into account  the noise effects associated with the dissipative dynamics and a realistic measurement process.  We sketch the main dynamics as follows, and details can be found in Appendix A.

Let $\hat{a}^{\prime}_i$ denote the canonical bosonic annihilation operator of the $i$th mode, $i=1,~2, \cdots,~N$. The $N\times N$ matrix $\tilde{H}[\epsilon]$ denotes the effective non-Hermitian Hamiltonian of $N$ resonant modes, where the parameter $\epsilon$ describes an infinitesimal change in the effective Hamiltonian $\tilde{H}[\epsilon]$. The aim of employing  non-Hermitian dynamics is to sense this infinitesimal change $\epsilon$.

Without loss of generality, we couple  the modes $i$ ($i=1,~2$) to a transmission line or a waveguide, respectively, through which a coherent drive with amplitude $\beta_i$ and frequency $w_{\textsf{dr},i}$ is injected. The coupling coefficient between mode $i$ and the corresponding waveguide is $k_i$, for $i=1,~2$.  We now assume that $w_{\textsf{dr},1}=w_{\textsf{dr},2}$. First, work in a rotating frame at the drive frequency  $w_{\textsf{dr},1}$, and let  $\hat{a}_i=\hat{a}^{\prime}_i e^{iw_{\textsf{dr},1}t}$. Then choose a frequency reference such that the real part of  ${\tilde{H}_{11}[0]}=0$. The full dynamics can be described by the  Heisenberg-Langevin equations \cite{Lau2018,Gardiner2000}:
\begin{equation}\label{main}
\begin{aligned}
\dot{\hat{a}}_i=&i\Delta \hat{a}_i-i\sum_{j}(\tilde{H}_{ij}[\epsilon]-i\frac{k_1}{2}\delta_{i1}\delta_{j1}-i\frac{k_2}{2}\delta_{i2}\delta_{j2})\hat{a}_{j}\\
&-i\delta_{i1}\sqrt{k_1}\beta_1-i\delta_{i2}\sqrt{k_2}\beta_2\\
&-i\delta_{i1}\sqrt{k_1}\hat{B}^{\textsf{in}}_1-i\delta_{i2}\sqrt{k_2}\hat{B}^{\textsf{in}}_2\\
&-i\sqrt{2}(\sum_{j=1}^{N_Y}Y_{ij}\hat{C}^{\textsf{in}\dagger}_{j}+\sum_{j=1}^{N_Z}Z_{ij}\hat{D}_{j}^{\textsf{in}}).
\end{aligned}
\end{equation}
Here, $\beta_i$ can be taken real and positive without loss of generality, and  $\Delta$ depicts the detuning of the drive frequency from the mode 1 resonance frequency $w_m$. The terms on the third line and fourth line  in Eq.~(\ref{main}) describe the noise effects. The noises $\hat{B}^{\textsf{in}}_{j}~(j=1,~2)$ denote the accompanied quantum noises of the coherent drives  coming from the input-output waveguides, whereas $\hat{C}^{\textsf{in}}_{j}$ ($\hat{D}^{\textsf{in}}_{j}$) are quantum noises arising from  dissipative baths depicting the gain (loss) processes with specific mode-bath coupling coefficients described by the matrix $Y$ ($Z$).

In order to ensure linearity and the Markovian  nature of the full dynamics,  $\hat{B}^{\textsf{in}}_{j}$, $\hat{C}^{\textsf{in}}_{j}$ and $\hat{D}^{\textsf{in}}_{j}$ are assumed to be quantum Gaussian white noises \cite{Gardiner2000}. We thus have
$$\langle Q(t)Q^\dagger(t')\rangle=(\bar{n}^{\textsf{th}}_Q+1)\delta(t-t'),$$ $$\langle Q^\dagger(t)Q(t')\rangle=\bar{n}^{\textsf{th}}_Q\delta(t-t'),$$ and $$\langle Q(t)Q(t')\rangle=0,$$ where $Q\in \{ \hat{B}^{\textsf{in}}_j,~\hat{C}^{\textsf{in}}_j,~\hat{D}^{\textsf{in}}_j \}$, and there are no correlations between different noise operators. The average $\langle \cdot \rangle$ represents the mean over the state of the bath degrees of freedom, and $\bar{n}^{\textsf{th}}_Q$ represents the average thermal occupancy of bath $Q$. In the following, we focus on the vacuum noise, i.e., $\bar{n}^{\textsf{th}}_Q=0$, while the formalism can be generalized to classical cases with $\bar{n}^{\textsf{th}}_Q\gg 1$.

\subsection{SNR and measurement rate}
As in \cite{Lau2018}, we take the standard  figure of merit  SNR or the equivalent measurement rate per photon, which determines the rate of SNR growing in time per photon to evaluate the sensitivity of measuring $\epsilon$. We start with specifying the homodyne measurement \cite{Wiseman2010,Moiseyev2011} which has been demonstrated being an optimal strategy if the driving field is sufficiently large \cite{Lau2018}.

The reflected field in the waveguide coupling to mode $1$ is described by $\hat{B}^{\textsf{out}}$, and from the standard input-output theory \cite{Gardiner2000} it obeys
\begin{equation*}\label{Bout}
\begin{aligned}
\hat{B}^{\textsf{out}}(t)=\beta_1+\hat{B}^{\textsf{in}}_1(t)-i\sqrt{k_1}\hat{a}_1(t),
\end{aligned}
\end{equation*}
where $\hat{B}^{\textsf{out}}(t)$ is related to $\hat{a}_1(t)$ with a dissipative rate $k_1$. Since we have assumed that the parameter change $\epsilon$ is small, the dependence of the mean value of the output field $\hat{B}^{\textsf{out}}(t)$
on $\epsilon$ is linear. Here, we  focus on the steady state values of the averages, by assuming that the measurement duration is sufficiently long such that any transient effects can be ignored or averaged out. We thus have
\begin{equation}\label{lambda}
\begin{aligned}
\langle \hat{B}^{\textsf{out}}\rangle_\epsilon \simeq \langle \hat{B}^{\textsf{out}}\rangle_0+\lambda \epsilon,
\end{aligned}
\end{equation}
where $\lambda$ is to be determined and possibly complex, and $\langle \cdot \rangle_\epsilon$ denotes the average calculated using Eq.~\eqref{main}, and  $\langle \cdot \rangle_0$ denotes the average calculated using Eq.~(\ref{main}) with $\epsilon=0$.

The homodyne detection is employed to extract the information of $\epsilon$ from the output field $\hat{B}^{\textsf{out}}(t)$. The homodyne current operator is
\begin{equation}\label{current}
\begin{aligned}
\hat{I}(t)\triangleq \sqrt{\frac{k_1}{2}}{\Big(}e^{i\phi} \hat{B}^{\textsf{out}}(t)+e^{-i\phi} \hat{B}^{\textsf{out}\dag}(t){\Big)}.
\end{aligned}
\end{equation}
We choose the phase $\phi$  as $\phi= -\arg \lambda$, following \cite{Lau2018}. This smart choice makes all the information of $\epsilon$ be contained in the real part of $e^{i\phi} \hat{B}^{\textsf{out}}$, and thus intuitively it is the best to measure the corresponding quadrature described by Eq.~(\ref{current}). In practice,  we prefer to integrate the homodyne current $\hat{I}(t)$ to average away the effects of noise, denoting this as $$\hat{m}(\tau)\triangleq \int^\tau_0 dt \hat{I}(t).$$

For the steady state averages in the long-$\tau$ limit, from Eqs.~(\ref{lambda}) and (\ref{current}), the \textit{signal power} of the small parameter change $\epsilon$ is
\begin{equation}\label{Signal}
\begin{aligned}
\mathcal {S}=[\langle \hat{m}(\tau)\rangle_\epsilon-\langle \hat{m}(\tau)\rangle_0]^2=2 k_1 \epsilon^2 |\lambda|^2 \tau^2.
\end{aligned}
\end{equation}

Since we are interested in an infinitesimal parameter change, as long as the non-Hermitian parameter change vanishes in the limit of $\epsilon\rightarrow0$, the \textit{noise power} of the integrated homodyne current in the long-time $\tau$ limit can be defined as $$\mathcal {N} \triangleq \langle \delta\hat{m}(\tau)\delta\hat{m}(\tau) \rangle_0,$$ where $$\delta\hat{m}=\hat{m}-\langle \hat{m} \rangle_0.$$

Note that what we are really interested in is the SNR, given a fixed number of photons used in the measurement. We thus define \textit{the measurement rate per photon} $\bar{\Gamma}_{\textsf{meas}}$, which quantifies the resolving power of weak continuous measurement in terms of the SNR per photon as
\begin{equation}\label{measurementrate}
\begin{aligned}
\frac{\mathcal {S}}{{\mathcal {N}}}\cdot\frac{1}{\bar{n}_{\textsf{tot}}}\triangleq\frac{\epsilon^2}{k_1^2} \tau\frac{ \Gamma_{\textsf{meas}}}{\bar{n}_{\textsf{tot}}}\triangleq\frac{\epsilon^2}{k_1^2} \tau\bar{\Gamma}_{\textsf{meas}},
\end{aligned}
\end{equation}
where $\bar{n}_{\textsf{tot}}\triangleq \sum_i \langle \hat{a}_i^\dagger \rangle \langle \hat{a}_i \rangle$ denotes  the total average photon number in all modes.  Note that we have neglected the incoherent photons injected by the bath. This is reasonable because on the one hand the injected incoherent photons are independent of the coherent drives; On the other hand, if the coherent drives are sufficiently large, the coherent drive-induced photons dominate the total photon number.  The appearance of the factor $k_1^{-2}$ in Eq.~(\ref{measurementrate}) is mainly for the the convenience of comparison with results in \cite{Lau2018} [see Eq. (16) therein]. From Eq.~(\ref{measurementrate}), it can be seen that  in the long-time limit, the SNR per photon grows  linearly with time $\tau$ in terms of the measurement rate per photon $\bar{\Gamma}_{\textsf{meas}}$.

\section{General bound of non-Hermitian sensing}
For stable non-Hermitian dynamics, we can derive a general limit on the measurement rate per photon or the corresponding SNR per photon whose details can be found in Appendix B.

Without loss of generality, we assume that the parameterized non-Hermitian Hamiltonian has the form
\begin{equation}\label{H1}
\begin{aligned}
\tilde{H}[\epsilon]=\tilde{H}[0]+\epsilon V,
\end{aligned}
\end{equation}
where $V$ is an $N\times N$ matrix which describes the coupling of the parameter $\epsilon$ to the unperturbed non-Hermitian Hamiltonian $\tilde{H}[0]$.

If the dynamics is stable, it is convenient to solve Eq.~(\ref{main}) in the frequency domain  in terms of the zero-frequency transfer matrix
\begin{equation*}\label{X}
\begin{aligned}
\tilde{\chi}^\Delta(\epsilon)\triangleq i k_1{\Big(}\Delta I-\tilde{H}[\epsilon]+i\frac{K_1}{2}+i\frac{K_2}{2}{\Big)}^{-1},
\end{aligned}
\end{equation*}
where $$K_1{_{ij}}=k_1 \delta_{i1}\delta_{j1}~ \text{and}~ K_2{_{ij}}=k_2 \delta_{i2}\delta_{j2}.$$ Define the unperturbed transfer matrix  $$\chi^\Delta\triangleq\tilde{\chi}^\Delta(0).$$ As demonstrated in \cite{Chen2019,Lau2018}, to obtain a full analysis of the sensitivity, not only the divergent eigenenergy susceptibility of $\tilde{H}[\epsilon]$ on $\epsilon$ should be considered, but also the left and right eigenvectors of $\tilde{H}[\epsilon]$ have to be taken into account, as the coalescence of the different eigenvectors may suppress or even cancel out the singular behavior of the divergence of the eigenenergy susceptibility.

From Eqs.~(\ref{main}) and (\ref{lambda}),  the linear response coefficient $\lambda$ can be derived as
\begin{equation}\label{lamda1}
\lambda=i \frac{\beta_1}{k_1}(\chi^{\Delta} V \chi^{\Delta} )_{11}+i \sqrt{\frac{k_2}{k_1}}\frac{\beta_2}{k_1}(\chi^{\Delta} V \chi^{\Delta} )_{12}.
\end{equation}
Combining Eqs.~(\ref{Signal}) and (\ref{lamda1}), the signal power becomes
\begin{equation}\label{Signal1}
\begin{aligned}
\mathcal {S}={2 k_1\epsilon^2 \tau^2}{\Big |}\frac{\beta_1}{k_1}(\chi^{\Delta} V \chi^{\Delta} )_{11}+\sqrt{\frac{k_2}{k_1}}\frac{\beta_2}{k_1}(\chi^{\Delta} V \chi^{\Delta} )_{12}{\Big|}^2.
\end{aligned}
\end{equation}

The noise power can be calculated from  the quantum Gaussian white noise properties as
\begin{equation}\label{Noise-2}
\begin{aligned}
\mathcal {N}=&\frac{k_1 \tau}{2}{\Big (}1+\frac{4}{k_1}(\chi^\Delta YY^\dagger \chi^{\Delta\dagger})_{11}{\Big)}.
\end{aligned}
\end{equation}
The first term here is the unavoidable shot noise, while the second term depends on how the gain processes are realized with coupling to dissipative baths.

We can optimize the mode-bath coupling matrices to minimize the noise power. The minimized noise power can be found as
\begin{equation}\label{mininoise}
\mathcal {N}_{\textsf{min}}=\frac{k_1 \tau}{2} {\big(}1+2\Xi\cdot\Theta(\Xi){\big)},
\end{equation}
where $$\Xi(\Delta)\triangleq-(\chi^\Delta_{11}+\chi^{\Delta\ast}_{11})+|\chi_{11}^{\Delta}|^2+\frac{k_2}{k_1}|\chi_{12}^{\Delta}|^2,$$ and $\Theta(\cdot)$ is the Heaviside step function. Note that our aim is to investigate the best possible measurement rate per photon in non-Hermitian sensing. In \cite{Lau2018}, it has been proved that for any fixed $\tilde{H}[0]$, one can always construct mode-bath couplings ($Y$ and $Z$) to attain the minimum noise power. A possible realization was also proposed, e.g., a two-mode non-reciprocal sensor can be implemented by coupling to an effective chiral waveguide which can be realized by using dynamic modulation and engineered dissipation \cite{Metelmann2015,Ozawa2019,ZYu2009,Ranzani2015,Sounas2017}.

From Eq.~(\ref{main}), the\textit{ total average photon number} in all modes induced by the coherent drives is
\begin{equation}\label{ntot}
\begin{aligned}
\bar{n}_{\textsf{tot}}=&\frac{1}{k_1} \beta_1^2 (\chi^{\Delta \dagger}\chi^\Delta)_{11}+\frac{k_2}{k_1^2} \beta_2^2 (\chi^{\Delta \dagger}\chi^\Delta)_{22}\\
&+\frac{\sqrt{k_1 k_2}}{k_1^2} \beta_1 \beta_2 {\Big(}(\chi^{\Delta \dagger}\chi^\Delta)_{12}+(\chi^{\Delta \dagger}\chi^\Delta)_{21}{\Big)}.
\end{aligned}
\end{equation}
By combining Eqs.~(\ref{measurementrate}) with  (\ref{Signal1})-(\ref{ntot}), we obtain  a general bound for the measurement rate per photon:
\begin{equation}\label{measurementoptimal}
\bar{\Gamma}_{\textsf{meas}}\leq\bar{\Gamma}_{\textsf{opt}}=\frac{k_1^2}{\epsilon^2}\cdot\frac{\mathcal {S}}{{\mathcal {N}_{\textsf{min}}}}\frac{1}{\tau}\cdot\frac{1}{\bar{n}_{\textsf{tot}}}.
\end{equation}
With this fundamental bound $\bar{\Gamma}_{\textsf{opt}}$  one can compare the best possible performance of sensors with different  non-Hermitian Hamiltonians.
\section{Two-mode non-Hermitian sensors}
Now we apply the  general bound (\ref{measurementoptimal}) on several typical kinds of two-mode systems which have been extensively studied in the context of EP sensors \cite{Sunada2017,Ren2017,Hodaei2017}. However, we stress that our results have nothing to do with EP.

Suppose that the coupling matrix in Eq.~(\ref{H1}) is $V=\frac{1}{2}\sigma_x,$ with $\sigma_x$ being the usual Pauli matrix  $\left(
                                                                                                                    \begin{array}{cc}
                                                                                                                      0 & 1 \\
                                                                                                                      1 & 0 \\
                                                                                                                    \end{array}
                                                                                                                  \right)$.
The best possible measurement rate per photon of two-mode sensors $\bar{\Gamma}_\textsf{{2-opt}}$  can be straightforwardly  calculated  from Eqs.~(\ref{Signal1}) -(\ref{measurementoptimal}) as

\begin{widetext}
\begin{equation}\label{bound1}
\begin{aligned}
\frac{|\chi^\Delta_{11}|^2|\chi^\Delta_{12}+\chi^\Delta_{21}|^2+2\sqrt{\eta}p\Re\{\chi^\Delta_{11}(\chi^\Delta_{12}+\chi^\Delta_{21})(\chi^{\Delta \ast 2}_{12}+\chi^{\Delta\ast}_{11}\chi^{\Delta\ast}_{22})\}+\eta p^2 |\chi^{\Delta 2}_{12}+\chi^\Delta_{11}\chi^\Delta_{22}|^2}{|\chi^\Delta_{11}|^2+|\chi^\Delta_{21}|^2+2\sqrt{\eta}p \Re \{ \chi^\Delta_{12}\chi^{\Delta \ast}_{11}+\chi^{\Delta \ast}_{21}\chi^\Delta_{22} \}+\eta p^2 (|\chi^\Delta_{12}|^2+|\chi^\Delta_{22}|^2)} \frac{k_1}{1+2\Xi\Theta(\Xi)},
\end{aligned}
\end{equation}
\end{widetext}
where  $p=\frac{\beta_2}{\beta1}$ is the ratio of the amplitudes of the coherent drives, and $\eta=\frac{k_2}{k_1}$.

If there is only one drive involved as the case in \cite{Lau2018}, then $p=0$ and $\eta=0$. From Eq.~(\ref{bound1}) we can obtain the main result in \cite{Lau2018} [see Eq. (27) therein].  However, if there are \textit{two} coherent drives, the amplitude ratio $p$ can be made arbitrarily large and  is independent of $\chi_{ij}^\Delta$.  Thus, from Eq.~(\ref{bound1}) we have
\begin{equation}\label{limitBound1}
\begin{aligned}
\bar{\Gamma}_\textsf{{2-opt}}\rightarrow \frac{|\chi^{\Delta 2}_{12}+\chi^\Delta_{11}\chi^\Delta_{22}|^2}{|\chi^\Delta_{12}|^2+|\chi^\Delta_{22}|^2}\cdot\frac{k_1}{1+2\Xi\Theta(\Xi)},
~\text{as}~p\rightarrow \infty.
\end{aligned}
\end{equation}

Note that if the non-Hermitian sensor is reciprocal, the magnitudes of the coupling between the two modes are the same, i.e., $|\tilde{H}_{12}|=|\tilde{H}_{21}|$, and this implies that $|\chi^\Delta_{12}|=|\chi^\Delta_{21}|$. For non-reciprocal sensors, $|\tilde{H}_{12}|\neq|\tilde{H}_{21}|$ and $|\chi^\Delta_{12}|\neq|\chi^\Delta_{21}|$, accordingly. From Eq.~\eqref{limitBound1}, it can be seen that there is no term $\chi^\Delta_{21}$ involved.  Thus, as long as the amplitude ratio $p$ is sufficiently large, a unified form of the bound on the best possible measurement rate per photon can be derived no matter whether the non-Hermitian sensor is reciprocal or non-reciprocal.

The bound in Eq.~(\ref{limitBound1}) can be further simplified under the condition   $|\chi_{12}^{\Delta}|\gg \max\{|\chi_{11}^{\Delta}|,~ |\chi_{22}^{\Delta}|,~1\}$. This can be seen as
\begin{equation*}\label{limitBound2}
\begin{split}
&~~~~\frac{|\chi^{\Delta 2}_{12}+\chi^\Delta_{11}\chi^\Delta_{22}|^2}{|\chi^\Delta_{12}|^2+|\chi^\Delta_{22}|^2}\cdot\frac{k_1}{1+2\Xi\Theta(\Xi)}\\
&=\frac{|\chi^{\Delta 2}_{12}+\chi^\Delta_{11}\chi^\Delta_{22}|^2}{|\chi^\Delta_{12}|^2+|\chi^\Delta_{22}|^2}\cdot \frac{k_1}{1+2(-2\Re\{\chi^\Delta_{11}\}+|\chi_{11}^{\Delta}|^2+\eta|\chi_{12}^{\Delta}|^2)}\\
&\rightarrow  \frac{1}{2}\cdot \frac{1}{\eta}\cdot k_1,~~\text{as}~~ |\chi_{12}^\Delta|\rightarrow \infty.
\end{split}
\end{equation*}

In practice, to approximately attain the uniform bound
\begin{equation}\label{bound-a}
\bar{\Gamma}_\textsf{{2-opt}}=\frac{1}{2}\cdot\frac{k_1}{k_2}\cdot k_1
\end{equation}
for non-Hermitian sensing,  one can first choose physical parameters such that
\begin{equation}\label{cond1-a}
|\chi_{12}^{\Delta}|\gg \max\{|\chi_{11}^{\Delta}|,~ |\chi_{22}^{\Delta}|,~1\}
\end{equation}
 holds. Then choose $p$ such that
\begin{equation}\label{cond2-a}
p \gg |\chi_{12}^\Delta|^3(|\chi_{12}^{\Delta}|+|\chi_{21}^\Delta|)
\end{equation}
to ensure the limit $p\rightarrow \infty$ in Eq.~(\ref{limitBound1}) being valid.

It can be seen that $\bar{\Gamma}_\textsf{{2-opt}}$  in Eq.~(\ref{bound-a}) depending  on  the ratio of the coupling coefficients $k_1$ and $k_2$,  can, in principle, be made arbitrarily large. This is quite different from the results in \cite{Lau2018}, where the measurement rate of reciprocal systems with only one coherent drive is fundamentally bounded. In this sense, we demonstrate that \textit{non-reciprocal sensors}, which are viewed as  powerful resources for quantum sensing, \textit{can be simulated by conventional reciprocal sensors with two coherent drives}. In practice, reciprocal sensors may be much easier to implement than non-reciprocal sensors.

Now let us first account for the condition (\ref{cond2-a}). Note that the optimal measurement rate in \cite{Lau2018} is given under the situation where the measurement noise is at the shot noise level. However, it can be calculated that in attaining $\bar{\Gamma}_\textsf{{2-opt}}$, the measurement noise is no longer shot noise, but contains noise  emanating  from the coupling between the coherent drive and mode $2$. Thus, to ensure that $\bar{\Gamma}_\textsf{{2-opt}}$ exceeds the optimal bound in \cite{Lau2018}, on the one hand  the ratio $k_1/k_2$ should be large. On the other hand,  the coupling coefficient  $k_2$ should be small while  $p=\frac{\beta_2}{\beta_1}$  sufficiently large, so that little noise is introduced through mode $2$, while the excitation signals through mode $2$ dominate in the total signal power.

To see  how the condition (\ref{cond1-a}) relates to physical parameters, such as the detuning, dissipative rates and coupling coefficients explicitly, we consider typical non-Hermitian mode systems which have been studied extensively in the literature.

\subsection{Reciprocal case}
First, we consider a reciprocal system in the form
\begin{equation}\label{Hexample}
\begin{aligned}
\tilde{H}_{\textsf{recip}}[0]=
\begin{pmatrix}
      -i\frac{\gamma_1}{2}& J \\
     J & -i\frac{\gamma_2}{2}
\end{pmatrix},
\end{aligned}
\end{equation}
where $J$ is the Hermitian coupling between the modes, while $\gamma_i~(i=1,~2)$ describe the possible gain/loss processes (depending on the sign) acting locally on each mode. It can be verified that if
\begin{equation}\label{r1}
\Delta\approx0,~k_i+\gamma_i\approx0,
\end{equation} and
\begin{equation}\label{r2}
 k_1 \gg |J| \gg \text{max}\{|\Delta|,~ |k_i+\gamma_i|\}~(\text{for}~ i=1,~ 2),
\end{equation}
 the condition (\ref{cond1-a})  holds.
Details can be found in Appendix C. This  corresponds to a setup where the coherent drives are resonant with the linear modes, the Hermitian coupling $J$ between the modes is relatively weak, and  the two modes are locally subject to gain, where $\gamma_i<0$ and $\gamma_i\approx -k_i$.

\begin{figure}[!htb]
  \centering
  \includegraphics[width=\hsize]{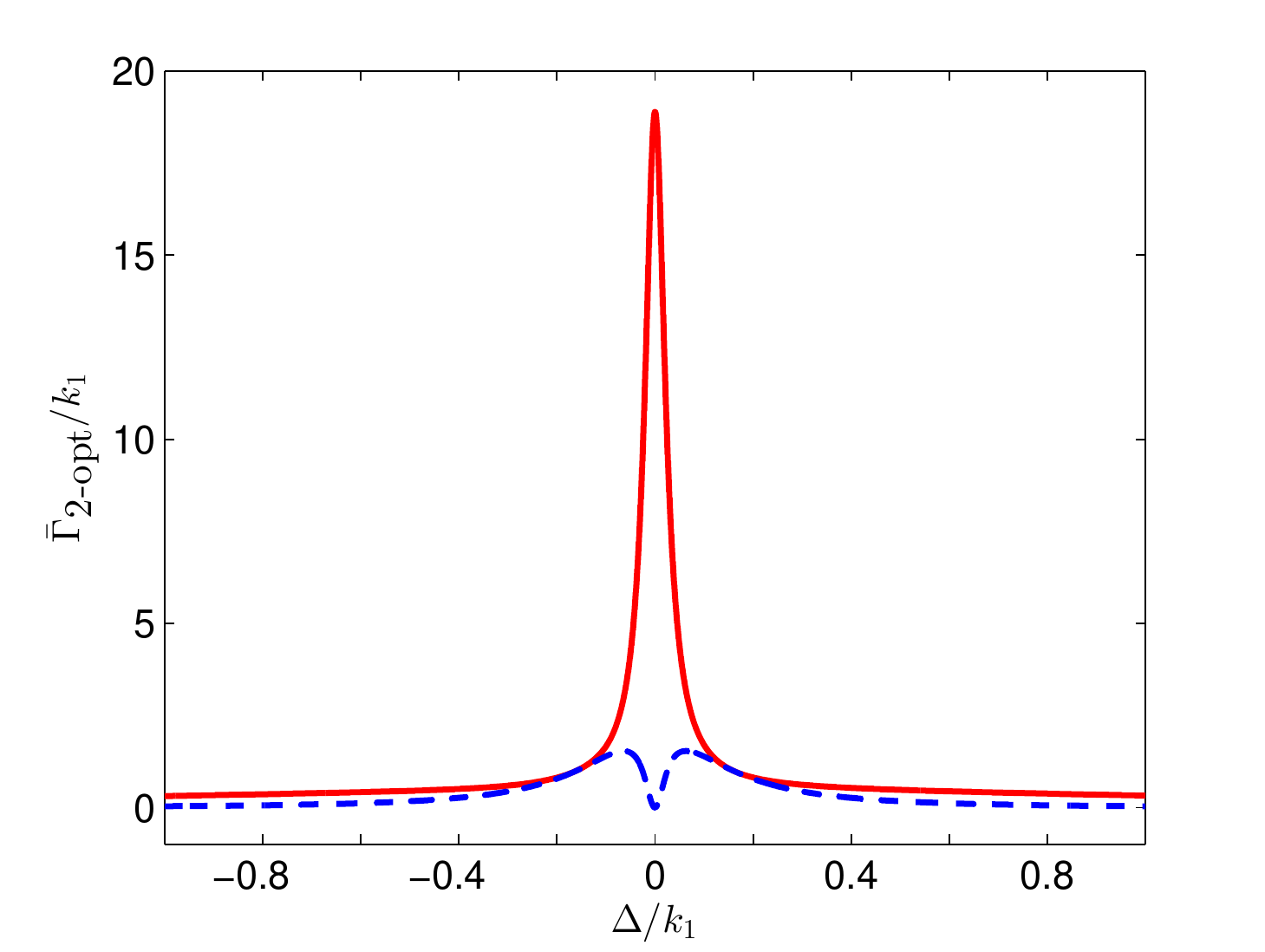}
  \caption{(Color online) The measurement rate per photon  ${\bar{\Gamma}_{\textsf{2-opt}}}/k_1$ versus the detuning $\Delta/k_1$. Blue dashed: one-drive reciprocal system with  gain, described by Eq.~(\ref{Hexample}) with $k_2=0$, $\gamma_1=-0.99k_1$, $\gamma_2=-0.011k_1$ and $J=0.16k_1$. Red solid: two-drive reciprocal system with gain,  described by Eq.~(\ref{Hexample}) with $k_2=0.01 k_1$, $\gamma_1=-0.99k_1$, $\gamma_2=-0.011k_1$, $J=0.16k_1$ and $\beta_2/\beta_1=30$. }
  \label{fig1}
\end{figure}

A concrete example is illustrated in Fig.~2 for  ${\bar{\Gamma}_{\textsf{2-opt}}}/k_1$, which, around the resonant frequency,   exceeds the fundamental limit   for reciprocal systems with only a single drive in \cite{Lau2018}. However,  for a fixed non-Hermitian system with $\gamma_1=0$, $\gamma_2=0.2k_1$ and $J=0.2k_1$,  no matter how $k_2$ and $\beta_2/\beta_1$ are adjusted, the performance of the measurement rate with only \textit{one} drive cannot be improved by \textit{two} excitation drives. The main reason is that amplification or gain from a local bath is a necessary ingredient for amplifying the signal power in the reciprocal case \cite{Lau2018}. If there is no gain, only with additional coherent drives, no enhancement can be achieved. The details can be found in Appendix C.
\subsection{Fully non-reciprocal case}
Now consider a fully non-reciprocal Hamiltonian
\begin{equation}\label{non-recip}
\begin{aligned}
\tilde{H}_{\textsf{non-reci}}[0]=
\begin{pmatrix}
      -i\frac{\gamma_1}{2}& J \\
     0 & \nu_2-i\frac{\gamma_2}{2}
\end{pmatrix},
\end{aligned}
\end{equation}
where $\nu_2$ is the frequency detuning of the two modes, and $J$ quantifies the non-reciprocal mode-mode coupling. It can be verified that, as long as the non-reciprocal coupling $|J|$ is sufficiently large, then condition (\ref{cond1-a}) holds.  Thus the amplification or gain from the local bath is not a necessary ingredient for non-reciprocal sensors.
\begin{figure}[!htb]
\centering
\includegraphics[width=\hsize]{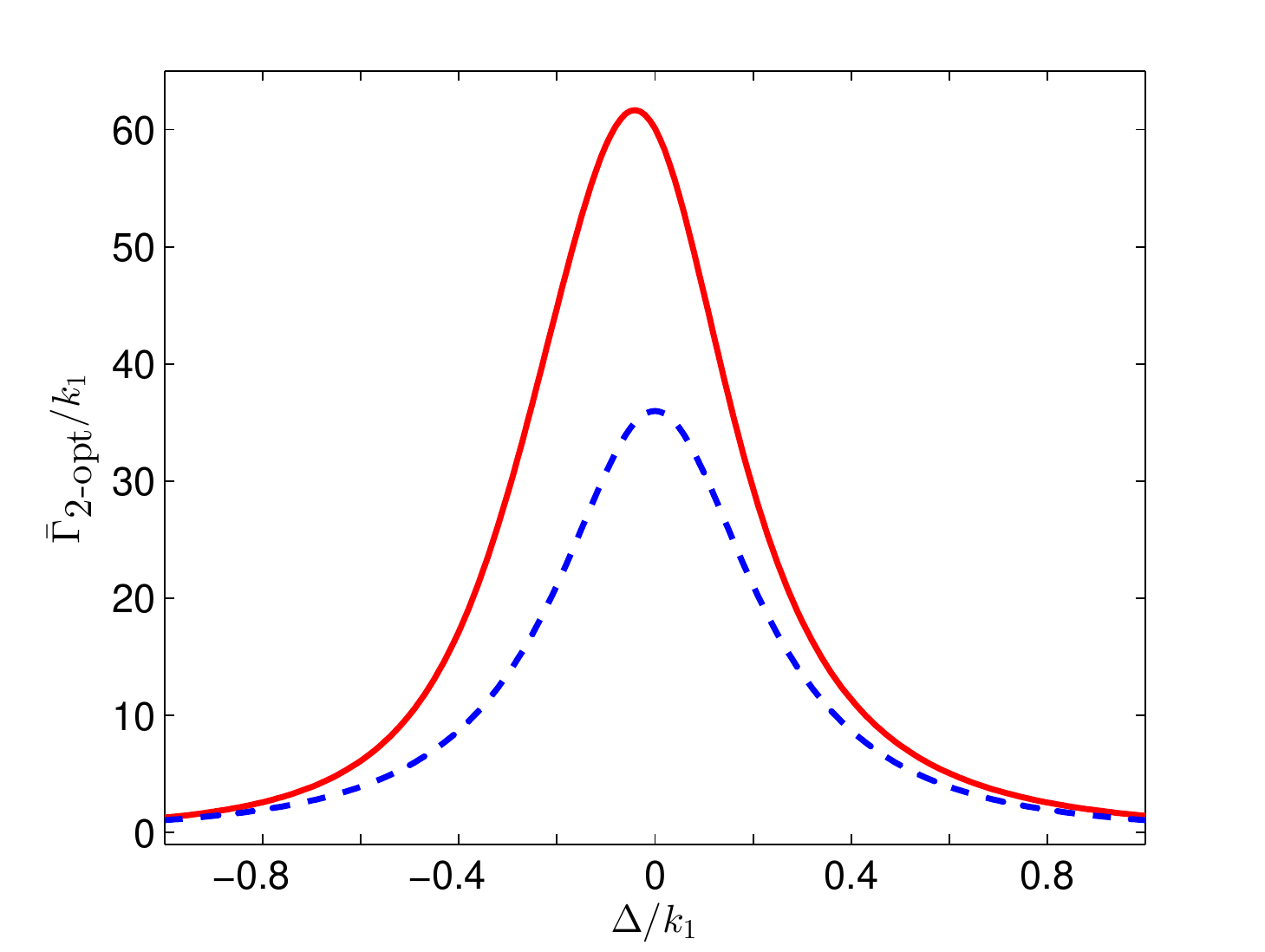}
\caption{(Color online) The measurement rate per photon  $\bar{\Gamma}_{\textsf{2-opt}}/k_1$. Blue dashed: one-drive non-reciprocal system described by Eq.~(\ref{non-recip}) with $k_2=0$, $\gamma_1=k_1$, $\gamma_2=0.5k_1$, $\nu_2=0$ and $J=1.5k_1$. Red solid: two-drive non-reciprocal system described by Eq.~(\ref{non-recip}) with $k_2=0.001 k_1$, $\gamma_1=k_1$, $\gamma_2=0.5k_1$, $\nu_2=0$, $J=1.5k_1$ and $\beta_2/\beta_1=5$. }
\label{fig1}
\end{figure}

As an illustration, we specialize the system with parameters  $\gamma_1=k_1,$ $\gamma_2=0.5k_1,~\nu_2=0$ and $J=1.5k_1$. Note that there is no coupling to gain baths. Let  $k_2/k_1=0.001$ and $\beta_2/\beta_1=5$. The measurement rates per photon are shown in Fig. 3.  It can be seen that in contrast to the reciprocal case, with two coherent drives the performance of the measurement rate per photon can be greatly improved for the non-reciprocal sensor even though there is no amplification or gain from the bath. The improvement is due to the fact that signals from the mode $2$ dominate in the total signal power.

Combined with the reciprocal case, it is worth pointing out that although the best possible measurement rate per photon is limited by the same bound as in Eq.~(\ref{bound-a}), the parameters (e.g., $J$) attaining this bound are quite different in reciprocal and non-reciprocal sensors.  In the reciprocal case, the physical parameters should satisfy conditions (\ref{r1}) and (\ref{r2}), while in the non-reciprocal case  one just ensures $|J|$  to be sufficiently large, which allows  more degrees of freedom for other parameters. Moreover, from Figs. 2 and 3, one can see that ${\bar{\Gamma}_{2\textsf{-opt}}}/k_1$ decreases more quickly as the detuning $\Delta$ deviates from 0 in the reciprocal case than that in the non-reciprocal case. Thus we claim that non-reciprocity  provides advantages for quantum sensing.

There have been several ways of realizing non-reciprocal interactions, ranging from photonic setups \cite{Lira2012,Tzuang2014,Eggleton2014}, optomechanical systems \cite{Fang2017,Bernier2017}, to classical microwave \cite{FangK2013,Estep2014} and superconducting circuits \cite{Abdo2013,Sliwa2015}. Although  these experiments were designed to build circulators and isolators, such systems could be exploited for enhanced sensing \cite{Lau2018}.

\section{Measurement rate with different drive frequencies}
In the above analysis, we have assumed that the frequencies of the two coherent  drives satisfy
$w_{\textsf{dr},1}= w_{\textsf{dr},2}$.  Now we consider how the coherent drive frequencies affect the best possible performance of the  measurement rate per photon. To be specific, we focus on the following  two typical cases where
$$|w_{\textsf{dr},1}- w_{\textsf{dr},2}|\gg |\Delta_i| $$
for $i=1$ and $2$, respectively. Here, $\Delta_i=w_{\textsf{dr},i}-w_m$ represents the detuning of the drive frequency $w_{\textsf{dr},i}$ from the mode 1 resonance frequency $w_m$.

First, suppose that   $|w_{\textsf{dr},1}- w_{\textsf{dr},2}|\gg |\Delta_1| $.
In this case, it is convenient to first work in a rotating frame at the drive frequency $w_{\textsf{dr},1}$, and then choose a frequency reference such that the real part of  ${\tilde{H}_{11}[0]}=0$. The Heisenberg-Langevin equations  become:
\begin{equation}\label{main1}
\begin{aligned}
\dot{\hat{a}}_i\!=&i\Delta_1 \hat{a}_i\!-\!i\sum_{j}(\tilde{H}_{ij}[\epsilon]\!-\!i\frac{k_1}{2}\delta_{i1}\delta_{j1}\!-\!i\frac{k_2}{2}\delta_{i2}\delta_{j2})\hat{a}_{j}\\
&-i\delta_{i1}\sqrt{k_1}\beta_1-i\delta_{i2}\sqrt{k_2}\beta_2e^{i(w_{\textsf{dr},1}-w_{\textsf{dr},2})t}\\
&-i\delta_{i1}\sqrt{k_1}\hat{B}^{\textsf{in}}_1-i\delta_{i2}\sqrt{k_2}\hat{B}^{\textsf{in}}_2\\
&-i\sqrt{2}(\sum_{j=1}^{N_Y}Y_{ij}\hat{C}^{\textsf{in}\dagger}_{j}+\sum_{j=1}^{N_Z}Z_{ij}\hat{D}_{j}^{\textsf{in}}).
\end{aligned}
\end{equation}

It can be seen that under the following condition,
 $$|w_{\textsf{dr},1}- w_{\textsf{dr},2}|\gg \max\{|\Delta_1|,~\|\tilde{H}\|,~k_i,\sqrt{k_i} \beta_i\}~(\text{for}~i=1,~2), $$
the rapid oscillation signal $\beta_2\exp\{i(w_{\textsf{dr},1}-w_{\textsf{dr},2})t\}$ from  drive $2$ can be averaged out in the long-time average limit due to the rotating-wave approximation (RWA).  This leads to (see Appendix D)
\begin{equation*}
\bar{\Gamma}_{\textsf{meas}}\leq \ \frac{4|(\chi^{\Delta_1} V \chi^{\Delta_1})_{11}|^2}{(\chi^{\Delta_1 \dagger}\chi^{\Delta_1})_{11}}\cdot k_1,
\end{equation*}
which does not exceed (and in general is smaller than) the fundamental bound in \cite{Lau2018}. This is because in this situation (although the rapid oscillation signal from drive 2 has been averaged out owing to the RWA)  the injected photons through mode $2$ still contribute to the total number of photons, and  the unavoidable accompanied  quantum noise  remains in the noise power.

Let us turn to the second case where $|w_{\textsf{dr},1}- w_{\textsf{dr},2}|\gg |\Delta_2| $.  Now it is convenient to  work in a rotating frame at the drive frequency $w_{\textsf{dr},2}$, and choose a frequency reference such that the real part of  ${\tilde{H}_{11}[0]}=0$, then the corresponding Heisenberg-Langevin equations become:
\begin{equation}\label{main3}
\begin{aligned}
\dot{\hat{a}}_i\!=&i\Delta_2 \hat{a}_i\!-\!i\sum_{j}(\tilde{H}_{ij}[\epsilon]\!-\!i\frac{k_1}{2}\delta_{i1}\delta_{j1}\!-\!i\frac{k_2}{2}\delta_{i2}\delta_{j2})\hat{a}_{j}\\
&-i\delta_{i1}\sqrt{k_1}\beta_1e^{i(w_{\textsf{dr},2}-w_{\textsf{dr},1})t}-i\delta_{i2}\sqrt{k_2}\beta_2\\
&-i\delta_{i1}\sqrt{k_1}\hat{B}^{\textsf{in}}_1-i\delta_{i2}\sqrt{k_2}\hat{B}^{\textsf{in}}_2\\
&-i\sqrt{2}(\sum_{j=1}^{N_Y}Y_{ij}\hat{C}^{\textsf{in}\dagger}_{j}+\sum_{j=1}^{N_Z}Z_{ij}\hat{D}_{j}^{\textsf{in}}).
\end{aligned}
\end{equation}

Similarly, under the RWA condition
$$|w_{\textsf{dr},1}- w_{\textsf{dr},2}|\gg \max\{|\Delta_2|,~\|\tilde{H}\|,~k_i,\sqrt{k_i} \beta_i\}~(\text{for}~i=1,~2),$$
the rapid oscillation excitation $\beta_1\exp\{i(w_{\textsf{dr},2}-w_{\textsf{dr},1})t\}$ from drive 1 can be averaged out.  However, it can be found in Appendix D that if  the conditions (\ref{cond1-a}) and (\ref{cond2-a}) hold with $\Delta$ being replaced by $\Delta_2$, the same bound as Eq. (\ref{bound-a})  can be established for two-mode non-Hermitian systems as
\begin{equation*}
\bar{\Gamma}_\textsf{{2-opt}}=\frac{1}{2}\cdot\frac{k_1}{k_2}\cdot k_1.
\end{equation*}
This is the situation where the excitation from mode 2 dominates in the total signal power. More importantly, the resulting signal power gain prevails in the competition with the noise power enhancement induced by the unavoidable associated noise. To sum up,  when utilizing two coherent drives whose frequencies are quite different, it is better to adjust the frequency of drive 2 to be near the resonance frequency of mode 1 .

\section{Conclusion and Discussion}
We have established a uniform bound for the best possible SNR or measurement rate per photon for reciprocal and non-reciprocal non-Hermitian quantum sensing with two coherent drives. The bound is only related to the coupling coefficients between the modes and coherent drives, and it constrains sensors no matter whether they are at EP or not.  The bound can be made arbitrarily large in principle and is approximately attainable. Our results highlight how the coherent excitation drives affect the SNR and show that with two drives conventional reciprocal sensors, which are easy to implement with current technology,  can simulate non-reciprocal sensors and enhance the performance of sensing.
%The results also demonstrate that  if more than two excitation signals can be injected into the modes through waveguides, one can simplify the setup with only two drives. One drive is applied to the mode, say mode 1, which is to be measured. Another drive is injected into the mode, which has the smallest coupling with the coherent drives.

\section*{Acknowledgments}
We thank the anonymous reviewers for their detailed and valuable comments. B. Q. acknowledged the support of National Natural Science Foundation of China (Nos. 61773370, 11688101 and 61833010), and D. D. acknowledged the partial support by the Australian Research Councils Discovery Projects funding scheme under Project DP190101566, the Alexander von Humboldt Foundation of Germany and the U.S. Office of Naval Research Global under Grant N62909-19-1-2129. F.N. is supported in part by: NTT Research,
Japan Science and Technology Agency (JST)
(via the Q-LEAP program, Moonshot R\&D Grant No. JPMJMS2061, and the CREST Grant No. JPMJCR1676),
Japan Society for the Promotion of Science (JSPS) (via the KAKENHI Grant No. JP20H00134 and the JSPS-RFBR Grant No. JPJSBP120194828),
Army Research Office (ARO) (Grant No. W911NF-18-1-0358),
Asian Office of Aerospace Research and Development (AOARD) (via Grant No. FA2386-20-1-4069),
and the Foundational Questions Institute Fund (FQXi) via Grant No. FQXi-IAF19-06.

\section*{Appendices}
In the appendices we demonstrate the detailed description of the non-Hermitian setup and derivations of the bound on the best possible measurement rate per photon.  The appendices are organized as follows. In Appendix A we describe  in detail the terms in the Heisenberg-Langevin equations  which depict the dynamics of the full non-Hermitian setup. In Appendix B we present the derivations of the signal power, noise power and the measurement rate per photon. We then consider the two-mode non-Hermitian systems in Appendix C. In Appendix D we consider the case where the frequencies of the two coherent drives are different and discuss how to obtain a better SNR by working in an appropriate rotating frame.

\appendix
\section{General non-Hermitian setup}

To make the paper self-contained, we describe the details of the non-Hermitian linear coupled modes in this section.

In many works, the dynamics of \emph{N} resonant modes is described by the linear equations:
\begin{equation*}
\begin{aligned}
\dot{\tilde{\alpha}}_{i}(t)=-i \omega_m \tilde{\alpha}_{i}(t) -i\sum_{j}\tilde{H}_{ij}[\epsilon]\tilde{\alpha}_{j}(t),
\end{aligned}
\end{equation*}
where $\tilde{\alpha}_{j}(t)$ is the amplitude of mode $j$, $\omega_m$ is the mode 1 resonance frequency, and the $N\times N$ matrix $\tilde{H}$ denotes an effective non-Hermitian Hamiltonian depicting both coherent and dissipative dynamics.  The parameter $\epsilon$ in the non-Hermitian Hamiltonian describes an infinitesimal perturbation, and our aim is to  sense this infinitesimal change.

To measure the perturbation, a general idea is to couple one of the modes, say mode 1, to an input-output waveguide. On the one hand we can use this port to excite the system with a coherent drive. On the other hand the reflected signal can be measured to estimate $\epsilon$. Unlike previous studies with one drive, we also couple mode 2 to another waveguide through which only the excitation signal is injected  but without measurement.
Coupling to  the waveguide results in extra damping, and accordingly
$$\tilde{H}_{ij}\rightarrow \tilde{H}_{ij}-i(k_1/2)\delta_{i1}\delta_{j1}-i(k_2/2)\delta_{i2}\delta_{j2},$$
where $k_i~(i=1,~2)$ is the coupling rate between mode $i$ and the corresponding waveguide. Now the system with coherent drives is described by the coupling-mode equations
\begin{equation}\label{average}
\begin{aligned}
\dot{\tilde{\alpha}}_{i}(t)\!=\!&-i \omega_m \tilde{\alpha}_{i}(t)\\
&-i\sum_{j}(\tilde{H}_{ij}[\epsilon]\!-\!i\frac{k_1}{2}\delta_{i1}\delta_{j1}\!-\!i\frac{k_2}{2}\delta_{i2}\delta_{j2})\tilde{\alpha}_{j}(t)\\
&-i\delta_{i1}\sqrt{k_1}\beta_1 e^{-i\omega_{\textsf{dr},1}t}\!-\!i\delta_{i2}\sqrt{k_2}\beta_2 e^{-i\omega_{\textsf{dr},2}t},
\end{aligned}
\end{equation}
where $\beta_{i}$  ($\omega_{\textsf{dr},i}$) is the amplitude  (frequency) of the $i$th coherent drive for  $i=1,~2$.

Since the dissipative dynamics is encoded in the anti-Hermitian part of $\tilde{H}$ which can be described as
\begin{equation*}\label{Hermitian}
\begin{aligned}
\frac{1}{2i}(\tilde{H}-\tilde{H}^\dagger)\equiv YY^\dagger-ZZ^\dagger,
\end{aligned}
\end{equation*}
where the matrix $YY^\dagger$ describes gain processes and $ZZ^\dagger$ represents loss processes. Let $Y$  be an $N\times N_Y$ matrix, and $Z$  be an $N\times N_Z$ matrix, i.e., the non-Hermitian dynamics is generated by coupling to $N_Y+N_Z$ distinct baths with the corresponding coupling constants described by $Y$ and $Z$.

Equation~(\ref{average}) can be viewed as a noise-averaged dynamics. Now we  describe the whole dynamics including the consistent noise processes as the Heisenberg-Langevin equations:
\begin{equation}\label{H-L_equtaion}
\begin{aligned}
\dot{\hat{a}}^{\prime}_i=&-i \omega_m \hat{a}^{\prime}_{i}-i\sum_{j}(\tilde{H}_{ij}[\epsilon]-i\frac{k_1}{2}\delta_{i1}\delta_{j1}-i\frac{k_2}{2}\delta_{i2}\delta_{j2})\hat{a}^{\prime}_{j}\\&-i\delta_{i1}\sqrt{k_1}\beta_1 e^{-i\omega_{\textsf{dr},1}t}
-i\delta_{i2}\sqrt{k_2}\beta_2 e^{-i\omega_{\textsf{dr},2}t}\\&-i\delta_{i1}\sqrt{k_1}\hat{B}^{\textsf{in}}_1-i\delta_{i2}\sqrt{k_2}\hat{B}^{\textsf{in}}_2\\
&-i\sqrt{2}(\sum_{j=1}^{N_Y}Y_{ij}\hat{C}^{\textsf{in}\dagger}_{j}+\sum_{j=1}^{N_Z}Z_{ij}\hat{D}_{j}^{\textsf{in}}).
\end{aligned}
\end{equation}
Here $\hat{a}^{\prime}_{i}$ denotes the canonical bosonic annihilation operator of the $i$th mode, $i=1,~ 2,~ \cdots,~ N$. Note that the terms in the first two lines are in the same form as those in Eq.~(\ref{average}), while the terms in the last two lines describe zero-mean noise effects.
The quantum noises $\hat{B}^{\textsf{in}}_{j}~(j=1,~ 2)$  come from the input-output waveguide, whereas $\hat{C}^{\textsf{in}}_{j}$ ($\hat{D}^{\textsf{in}}_{j}$) are quantum noises arising from  dissipative baths used to realize the gain (loss) parts of the dynamics with specific mode-bath coupling coefficients described by the matrix $Y$ ($Z$). The quantum noises $\hat{B}^{\textsf{in}}_{j}$, $\hat{C}^{\textsf{in}}_{j}$ and $\hat{D}^{\textsf{in}}_{j}$ are assumed to be quantum Gaussian white noises, which satisfy
$$\langle Q(t)Q^\dagger(t')\rangle=(\bar{n}^{\textsf{th}}_Q+1)\delta(t-t'),$$ $$\langle Q^\dagger(t)Q(t')\rangle=\bar{n}^{\textsf{th}}_Q\delta(t-t'),$$  and $$\langle Q(t)Q(t')\rangle=0,$$
where $Q\in \{ \hat{B}^{\textsf{in}}_j,~ \hat{C}^{\textsf{in}}_j,~\hat{D}^{\textsf{in}}_j \},$ and the correlations between different noise operators vanish.

Assume  $w_{\textsf{dr},1}$=$w_{\textsf{dr},2}$. First, work in a rotating frame at the drive frequency  $w_{\textsf{dr},1}$, and let  $\hat{a}_i=\hat{a}^{\prime}_i e^{iw_{\textsf{dr},1}t}$. Then choose a frequency reference such that the real part of  ${\tilde{H}_{11}[0]}=0$.  The  Heisenberg-Langevin equations (\ref{H-L_equtaion}) become
\begin{equation}\label{main01}
\begin{aligned}
\dot{\hat{a}}_i\!=&i\Delta \hat{a}_i\!-\!i\sum_{j}(\tilde{H}_{ij}[\epsilon]\!-\!i\frac{k_1}{2}\delta_{i1}\delta_{j1}\!-\!i\frac{k_2}{2}\delta_{i2}\delta_{j2})\hat{a}_{j}\\
&-i\delta_{i1}\sqrt{k_1}\beta_1-i\delta_{i2}\sqrt{k_2}\beta_2\\
&-i\delta_{i1}\sqrt{k_1}\hat{B}^{\textsf{in}}_1-i\delta_{i2}\sqrt{k_2}\hat{B}^{\textsf{in}}_2\\
&-i\sqrt{2}(\sum_{j=1}^{N_Y}Y_{ij}\hat{C}^{\textsf{in}\dagger}_{j}+\sum_{j=1}^{N_Z}Z_{ij}\hat{D}_{j}^{\textsf{in}}),
\end{aligned}
\end{equation}
where  $\Delta$ represents the detuning of the drive frequency from the mode 1 resonance frequency. Here we still adopt the same symbols for the annihilation operators  and noises.

\section{Derivations of the SNR and measurement rate}
In this section, we present the detailed calculation in deriving the SNR and measurement rate per photon.

Without loss of generality we assume that the parameterized non-Hermitian Hamiltonian is in the form
\begin{equation*}\label{H11}
\begin{aligned}
\tilde{H}[\epsilon]=\tilde{H}[0]+\epsilon V,
\end{aligned}
\end{equation*}
where $V$ is an $N\times N$ matrix which describes the coupling of the parameter $\epsilon$ to the unperturbed non-Hermitian Hamiltonian $\tilde{H}[0]$. If the non-Hermitian dynamics (\ref{main01}) is stable, it is  convenient to transfer into the frequency domain to solve Eq.~(\ref{main01}) in terms of the zero-frequency transfer matrix
\begin{equation}\label{kappamatrix}
\begin{aligned}
\tilde{\chi}^\Delta(\epsilon)\triangleq i k_1 {\Big(}\Delta I-\tilde{H}[\epsilon]+i\frac{K_1}{2}+i\frac{K_2}{2}{\Big)}^{-1},
\end{aligned}
\end{equation}
where $$K_1{_{ij}}=k_1 \delta_{i1}\delta_{j1}~ \text{and}~ K_2{_{ij}}=k_2 \delta_{i2}\delta_{j2}.$$ Moreover, define  the unperturbed transfer matrix as $$\chi^\Delta\triangleq\tilde{\chi}^\Delta(0).$$

To be specific, from Eq.~(\ref{main}) or (\ref{main01}), if the time $t$ is sufficiently large, the annihilation operator of the mode $l$ can be described as
\begin{equation*}
\begin{aligned}
\hat{a}_l=&\frac{\sqrt{k_1}}{i k_1}\beta_1\tilde{\chi}_{l1}^{\Delta}+\frac{\sqrt{k_2}}{i k_1}\beta_2\tilde{\chi}_{l2}^{\Delta}\\
&+\frac{\sqrt{k_1}}{i k_1}\hat{B}^{\textsf{in}}_1\tilde{\chi}_{l1}^{\Delta}+\frac{\sqrt{k_2}}{i k_1}\hat{B}^{\textsf{in}}_2\tilde{\chi}_{l2}^{\Delta}\\
&+\frac{\sqrt{2}}{i k_1}{\big(}\sum^{N_Y}_{j=1}(\tilde{\chi}^\Delta Y)_{lj}C^{\textsf{in}\dagger}_j+\sum^{N_Z}_{j=1}(\tilde{\chi}^\Delta Z)_{lj}D^{\textsf{in}}_j{\big)}.
\end{aligned}
\end{equation*}
Thus the reflected field
\begin{equation*}\label{Bout1}
\begin{aligned}
\hat{B}^{\textsf{out}}=&\beta_1+\hat{B}^{\textsf{in}}_1(t)-i\sqrt{k_1}\hat{a}_1(t)\\
=&(1-\tilde{\chi}^\Delta_{11})\beta_1-\sqrt{\frac{k_2}{k_1}}\tilde{\chi}^\Delta_{12}\beta_2\\
&+(1-\tilde{\chi}^\Delta_{11})\hat{B}^{\textsf{in}}_1
-\sqrt{\frac{k_2}{k_1}}\tilde{\chi}^\Delta_{12}\hat{B}^{\textsf{in}}_2\\
&-\sqrt{\frac{2}{k_1}}{\big(}\sum^{N_Y}_{j=1}(\tilde{\chi}^\Delta Y)_{1j}C^{\textsf{in}\dagger}_j+\sum^{N_Z}_{j=1}(\tilde{\chi}^\Delta Z)_{1j}D^{\textsf{in}}_j{\big)}.
\end{aligned}
\end{equation*}
Since we are considering an infinitesimal perturbation $\epsilon$, the change of the mean of the reflected field can be represented as
\begin{equation*}\label{Bout2}
\begin{aligned}
&\langle \hat{B}^{\textsf{out}}\rangle_\epsilon - \langle \hat{B}^{\textsf{out}}\rangle_0\\
=&\frac{-1}{\sqrt{k_1}}{\big(}\tilde{\chi}_{11}^{\Delta}(\epsilon)-\tilde{\chi}_{11}^{\Delta}(0){\big)}\sqrt{k_1}\beta_1\\
&+\frac{-1}{\sqrt{k_1}}{\big(}\tilde{\chi}_{12}^{\Delta}(\epsilon)-\tilde{\chi}_{12}^{\Delta}(0){\big)}\sqrt{k_2}\beta_2\\
=&\lambda \epsilon,
\end{aligned}
\end{equation*}
where
\begin{equation*}
\begin{aligned}
\lambda&=-\beta_1\frac{d}{d\epsilon}\tilde{\chi}^{\Delta }_{11}(\epsilon)|_{\epsilon=0}-\sqrt{\frac{k_2}{k_1}}\beta_2\frac{d}{d\epsilon}
\tilde{\chi}^{\Delta }_{12}(\epsilon)|_{\epsilon=0}\\
&=i \frac{\beta_1}{k_1}(\chi^{\Delta} V \chi^{\Delta} )_{11}+i \sqrt{\frac{k_2}{k_1}}\frac{\beta_2}{k_1}(\chi^{\Delta} V \chi^{\Delta} )_{12}.
\end{aligned}
\end{equation*}

According to the definition of the signal power in  Eq.~(\ref{Signal}),
\begin{equation}\label{ASignal1}
\begin{aligned}
\mathcal {S}\!=\!{2 k_1\epsilon^2 \tau^2}{\Big |}\frac{\beta_1}{k_1}(\chi^{\Delta} V \chi^{\Delta} )_{11}\!+\!\sqrt{\frac{k_2}{k_1}}\frac{\beta_2}{k_1}(\chi^{\Delta} V \chi^{\Delta} )_{12}{\Big|}^2.
\end{aligned}
\end{equation}

The total photon number in all modes is described by
\begin{equation}\label{Antot}
\begin{aligned}
\bar{n}_{\textsf{tot}}=&\sum_i \langle \hat{a}_i^\dagger \rangle \langle \hat{a}_i \rangle\\
=&\frac{1}{k_1} \beta_1^2 (\chi^{\Delta \dagger}\chi^\Delta)_{11}+\frac{k_2}{k_1^2} \beta_2^2 (\chi^{\Delta \dagger}\chi^\Delta)_{22}\\
&+\frac{\sqrt{k_1 k_2}}{k_1^2} \beta_1 \beta_2 {\big(}(\chi^{\Delta \dagger}\chi^\Delta)_{12}+(\chi^{\Delta \dagger}\chi^\Delta)_{21}{\big)}.
\end{aligned}
\end{equation}

The noise power $\mathcal{N}$ is defined as $$\mathcal {N} \triangleq \langle \delta\hat{m}(\tau)\delta\hat{m}(\tau) \rangle_0,$$ where $$\delta\hat{m}(\tau)\triangleq\hat{m}(\tau)-\langle \hat{m}(\tau) \rangle_0,$$ and $$\hat{m}(\tau)\triangleq \int^\tau_0 dt \hat{I}(t).$$ Combining  Eqs.~(\ref{main})-(\ref{current}), we have
\begin{equation*}\label{delta-m}
\begin{aligned}
\delta \hat{m}=\sqrt{\frac{k_1}{2}}& \int^\tau_0 dt{\Big[}(1-\chi^\Delta_{11})\hat{B}^{\textsf{in}}+(1-\chi^{\Delta\ast}_{11})\hat{B}^{\textsf{in}\dagger}_1\\
&~-\sqrt{\frac{k_2}{k_1}}\chi^\Delta_{12}\hat{B}^{\textsf{in}}_2-\sqrt{\frac{k_2}{k_1}}\chi^{\Delta\ast}_{12}\hat{B}^{\textsf{\textsf{in}}\dagger}_2\\
&-\sqrt{\frac{2}{k_1}}\sum^{N_Y}_{j=1}{\big(}(\chi^\Delta Y)_{1j}C^{\textsf{in}\dagger}_j+(\chi^\Delta Y)_{1j}^\ast C_j^{\textsf{in}}{\big)}\\
&-\sqrt{\frac{2}{k_1}}\sum^{N_Z}_{j=1}{\big(}(\chi^\Delta Z)_{1j}D^{\textsf{in}}_j+(\chi^\Delta Z)_{1j}^\ast D^{\textsf{in}\dagger}_j{\big)}{\Big]},
\end{aligned}
\end{equation*}
where the phase $\phi$ in Eq.~(\ref{current}) has been included in the noise operators. Employing  the properties of quantum Gaussian white noise, the noise power is
\begin{equation}\label{Noise-1}
\begin{aligned}
\mathcal {N}=&\frac{k_1 \tau}{2}{\Big (}|1-\chi^\Delta_{11}|^2+\frac{k_2}{k_1}|\chi_{12}^\Delta|^2\\
&+\frac{2}{k_1}{\big(}(\chi^\Delta Y Y^\dagger \chi^{\Delta \dagger})_{11}+(\chi^\Delta Z Z^\dagger \chi^{\Delta \dagger})_{11}{\big)}{\Big)}.
\end{aligned}
\end{equation}
From Eq.~(\ref{kappamatrix}), we have
\begin{equation*}
\begin{aligned}
\chi^{\Delta -1}+(\chi^{\Delta \dagger})^{-1}=-\frac{2}{k_1}(YY^\dagger-ZZ^\dagger-\frac{1}{2}K_1-\frac{1}{2}K_2).
\end{aligned}
\end{equation*}
Thus,
\begin{equation}\label{eq}
\begin{aligned}
\chi^\Delta_{11}\!+\!\chi^{\Delta\ast}_{11}\!=\!-\frac{2}{k_1}{\big[}&(\chi^\Delta YY^\dagger\chi^{\Delta\dagger})_{11}\!-\!(\chi^\Delta ZZ^\dagger\chi^{\Delta\dagger})_{11}\\
&-\frac{1}{2}k_1|\chi^\Delta_{11}|^2-\frac{1}{2}k_2|\chi^\Delta_{12}|^2{\big]}.
\end{aligned}
\end{equation}
This  is combined with Eq.~(\ref{Noise-1}), and we have the noise power  as
\begin{equation}\label{Noise-21}
\begin{aligned}
\mathcal {N}=&\frac{k_1 \tau}{2} {\Big(}1+\frac{4}{k_1}(\chi^\Delta YY^\dagger \chi^{\Delta\dagger})_{11}{\Big)}.
\end{aligned}
\end{equation}
From Eqs.~(\ref{eq}), (\ref{Noise-21}) and the fact that $ZZ^{\dagger}$ is positive semidefinite, the minimized noise power can be found as
\begin{equation}\label{mininoise}
\mathcal {N}_{\textsf{min}}=\frac{k_1 \tau}{2} {\big(}1+2\Xi\cdot\Theta(\Xi){\big)},
\end{equation}
where $$\Xi(\Delta)\triangleq-(\chi^\Delta_{11}+\chi^{\Delta\ast}_{11})+|\chi_{11}^{\Delta}|^2+\frac{k_2}{k_1}|\chi_{12}^{\Delta}|^2,$$ and $\Theta(\cdot)$ is the Heaviside step function. It is worth stressing that for any fixed $\tilde{H}[0]$, one can always construct mode-bath couplings ($Y$ and $Z$) to attain the minimum possible noise power.

Now it is straightforward to combine Eqs.~(\ref{measurementrate}), (\ref{ASignal1}),  (\ref{Antot}), (\ref{Noise-21}) and (\ref{mininoise}) to derive a general bound for the measurement rate per photon as
\begin{widetext}
\begin{eqnarray}\label{bound0}
\begin{aligned}
\bar{\Gamma}_{\textsf{meas}}&=
\frac{k_1\cdot 4{\big|}\beta_1(\chi^{\Delta} V \chi^{\Delta} )_{11}+\sqrt{\frac{k_2}{k_1}}\beta_2(\chi^{\Delta} V \chi^{\Delta} )_{12}{\big|}^2}{{\big(}\beta_1^2 (\chi^{\Delta \dagger}\chi^\Delta)_{11}+\frac{k_2}{k_1} \beta_2^2 (\chi^{\Delta \dagger}\chi^\Delta)_{22}
+2\sqrt{\frac{k_2}{k_1} }\beta_1 \beta_2 \Re\{(\chi^{\Delta \dagger}\chi^\Delta)_{12}\}{\big)}{\big(}1+\frac{4}{k_1}(\chi^\Delta Y Y^\dagger \chi^{\Delta \dagger})_{11}{\big)}}\\
&\leq \frac{ k_1\cdot4{\big|}\beta_1(\chi^{\Delta} V \chi^{\Delta} )_{11}+\sqrt{\frac{k_2}{k_1}}\beta_2(\chi^{\Delta} V \chi^{\Delta} )_{12}{\big|}^2}{{\big(}\beta_1^2 (\chi^{\Delta \dagger}\chi^\Delta)_{11}+\frac{k_2}{k_1} \beta_2^2 (\chi^{\Delta \dagger}\chi^\Delta)_{22}
+2\sqrt{\frac{k_2}{k_1} }\beta_1 \beta_2 \Re\{(\chi^{\Delta \dagger}\chi^\Delta)_{12}\}{\big)}{\big(}1+2\Xi\cdot\Theta(\Xi){\big)}}\\
&\triangleq\bar{\Gamma}_{\textsf{opt}},
\end{aligned}
\end{eqnarray}
\end{widetext}
where $\Re\{\cdot \}$ denotes the real part of the variable.
With this fundamental bound we can compare the best possible performance of senors with different non-Hermitian Hamiltonians.

\section{ Two-mode non-Hermitian sensors  }

\subsection{Physical parameter conditions}

Now  consider the physical parameter conditions satisfying Eq.~(\ref{cond1-a}) for  two-mode reciprocal and non-reciprocal sensors, respectively.

First, consider a reciprocal system in the form of
\begin{equation*}
\begin{aligned}
\tilde{H}_{\textsf{recip}}[0]=
\begin{pmatrix}
      -i\frac{\gamma_1}{2}& J \\
     J & -i\frac{\gamma_2}{2}
\end{pmatrix},
\end{aligned}
\end{equation*}
where $J$ is the Hermitian coupling between the modes, while $\gamma_i~(i=1,~2)$ describe the possible gain/loss processes (depending on the sign) acting locally on each mode. The corresponding matrix $\chi^{\Delta}_{\textsf{recip}}$ is
\begin{equation*}
\begin{aligned}
\frac{i k_1}{\mathcal{G}}
\begin{pmatrix}
\Delta+\frac{i}{2}(k_2+\gamma_2) &  J \\
     J & \Delta+\frac{i}{2}(k_1+\gamma_1)
\end{pmatrix},
\end{aligned}
\end{equation*}
where $$\mathcal{G}=-J^2+(\Delta+\frac{i}{2}(k_1+\gamma_1))(\Delta+\frac{i}{2}(k_2+\gamma_2)).$$
It can be verified that if $$\Delta\approx0,~k_i+\gamma_i\approx0,$$ and
$$ k_1 \gg |J| \gg \text{max}\{|\Delta|,~ |k_i+\gamma_i|\}~(\text{for}~ i=1,~ 2),$$ the condition (\ref{cond1-a})  holds.

Similarly, consider a fully non-reciprocal system in the form
\begin{equation*}
\begin{aligned}
\tilde{H}_{\textsf{recip}}[0]=
\begin{pmatrix}
      -i\frac{\gamma_1}{2}& J \\
     0 & -i\frac{\gamma_2}{2}
\end{pmatrix}.
\end{aligned}
\end{equation*}
The corresponding matrix $\chi^{\Delta}_{\textsf{non-recip}}$ is
\begin{equation*}
\begin{aligned}
i k_1
\begin{pmatrix}
\frac{1}{\Delta+\frac{i}{2}(k_1+\gamma_1)} &  \frac{J}{(\Delta+\frac{i}{2}(k_1+\gamma_1))(\Delta+\frac{i}{2}(k_2+\gamma_2))} \\
     0 & \frac{1}{\Delta+\frac{i}{2}(k_2+\gamma_2)}
\end{pmatrix}.
\end{aligned}
\end{equation*}
It can be seen that the condition \eqref{cond1-a} can be ensured as long as  $|J|$ is sufficiently large.

\subsection{Case with no gain in reciprocal sensors}
In the paper, we have accounted for the uniform bound (\ref{bound-a}) and the  two conditions explicitly. Here we consider the case where there is no gain in the reciprocal process.  Consider a reciprocal system in the form
\begin{equation*}\label{example}
\begin{aligned}
\tilde{H}_{\textsf{recip}}[0]=
\begin{pmatrix}
      -i\frac{\gamma_1}{2}& J \\
     J & -i\frac{\gamma_2}{2}
\end{pmatrix}
\end{aligned}.
\end{equation*}
 If one  $\gamma_i$ satisfies $\gamma_i<0$, then there is a local gain from the bath.

Suppose that the reciprocal system has parameters $\gamma_1=0$, $\gamma_2=0.2k_1$ and $J=0.2k_1$. It can be verified  that with only a single drive, the measurement rate per photon $\bar{\Gamma}_{\textsf{opt,single}}/{k_1}$ is approximately equal to 5.67003.  We wonder that with an additional drive whether the performance of the measurement rate per photon can be improved.  Since  the measurement rate per photon $\bar{\Gamma}_{\textsf{opt}}/{k_1}$ in this case is a function of $\eta=k_2/k_1$, $p=\beta_2/\beta_1$ and the detuning $\Delta$, the question can be converted into whether the solution set of the following inequalities is empty:
$$ \left\{
\begin{aligned}
& \frac{\bar{\Gamma}_{\textsf{opt}}(\Delta,\eta,p)}{k_1 } \geq\mu, \\
& p\geq0, \\
& \eta\geq0,
\end{aligned}
\right.
$$
where $\mu$ is set to be 5.67003.
It can be verified that no matter how $\eta$, $p$ and $\Delta$ are adjusted, one cannot improve the performance of the measurement rate corresponding to the case where there is only a single drive on mode $1$. The main reason is that for reciprocal systems, amplification or gain from local bath is a necessary ingredient for amplifying the signal power. If there is no gain, only with additional coherent drives, no enhancement can be achieved.

\section{Different Drive Frequencies}

First, consider the case where $|w_{\textsf{dr},1}- w_{\textsf{dr},2}|\gg |\Delta_1| $.  It is convenient to work in a rotating frame at the drive frequency $w_{\textsf{dr},1}$.

From Eq.~(\ref{main1}), if we consider the long-time average, under the following RWA condition
$$|w_{\textsf{dr},1}- w_{\textsf{dr},2}|\gg \max\{|\Delta_1|,~\|\tilde{H}\|,~k_i,\sqrt{k_i} \beta_i \}~(\text{for}~i=1,~2), $$
the excitation containing the rapid oscillation $\sqrt{k_2}\beta_2\exp\{i(\omega_{\textsf{dr},1}-\omega_{\textsf{dr},2})t\}$ can be averaged out. Thus, for  sufficiently large $t$, the annihilation operator of the mode $l$ is described as
\begin{equation*}
\begin{aligned}
\hat{a}_l(t)=&\frac{\sqrt{k_1}}{i k_1}\beta_1\tilde{\chi}_{l1}^{\Delta_1}
+\frac{\sqrt{k_1}}{i k_1}\hat{B}^{\textsf{in}}_1\tilde{\chi}_{l1}^{\Delta_1}+\frac{\sqrt{k_2}}{i k_1}\hat{B}^{\textsf{in}}_2\tilde{\chi}_{l2}^{\Delta_1}\\
&+\frac{\sqrt{2}}{i k_1}{\big(}\sum^{N_Y}_{j=1}(\tilde{\chi}^{\Delta_1} Y)_{lj}C^{\textsf{in}\dagger}_j+\sum^{N_Z}_{j=1}(\tilde{\chi}^{\Delta_1} Z)_{lj}D^{\textsf{in}}_j{\big)},
\end{aligned}
\end{equation*}
where the transfer matrix $\tilde{\chi}^{\Delta}$ is defined in Eq.~(\ref{kappamatrix}).

 Based on this and following a similar analysis to that in Appendix B, the parameter $\lambda$ is described as
\begin{equation*}
\lambda=i \frac{\beta_1}{k_1}(\chi^{\Delta_1} V \chi^{\Delta_1} )_{11}.
\end{equation*}
Thus the signal power is
\begin{equation}\label{signal1}
\mathcal{S}=2\epsilon^2\tau^2\frac{\beta_1^2}{k_1}{\big|}(\chi^{\Delta_1} V \chi^{\Delta_1} )_{11}{\big|}^2.
\end{equation}
It only contains the  signals from  mode $1$, as the signals from mode $2$ have been averaged out in the long-time limit.

The total number of photons is
\begin{equation}\label{ntot1}
\bar{n}_{\textsf{tot}}
=\frac{\beta_1^2}{k_1} (\chi^{\Delta_1 \dagger}\chi^{\Delta_1})_{11}+\frac{k_2\beta_2^2 }{k_1^2} (\chi^{\Delta_1^\prime\dagger}\chi^{\Delta_1^\prime})_{22},
\end{equation}
where $\Delta_1^\prime=\Delta_1-\omega_{\textsf{dr},1}+\omega_{\textsf{dr},2}.$

In contrast to the signal power, the photons injected  through mode $2$ still contribute to the total number of photons. Thus the signal power per photon is reduced compared to the case where there is only one drive coupling with mode $1$.

Moreover, we have the noise power as
\begin{equation}\label{Noise-3}
\begin{aligned}
\mathcal {N}=&\frac{k_1 \tau}{2} {\Big(}1+\frac{4}{k_1}(\chi^{\Delta_1} YY^\dagger \chi^{\Delta_1\dagger})_{11}{\Big)}.
\end{aligned}
\end{equation}
From Eqs.~(\ref{signal1}), (\ref{ntot1}) and  (\ref{Noise-3}), the measurement rate per photon is described as
\begin{widetext}
\begin{eqnarray*}\label{bound}
\begin{aligned}
\bar{\Gamma}_{\textsf{meas}}&=k_1\cdot
\frac{4{\big|}(\chi^{\Delta_1} V \chi^{\Delta_1} )_{11}{\big|}^2}{{\big(}(\chi^{{\Delta_1} \dagger}\chi^{\Delta_1})_{11}+\frac{k_2}{k_1} \frac{\beta_2^2}{\beta_1^2} (\chi^{{\Delta_1^{\prime}} \dagger}\chi^{\Delta_1^{\prime}})_{22}{\big)}
{\big(}1+\frac{4}{k_1}(\chi^{\Delta_1} Y Y^\dagger \chi^{{\Delta_1} \dagger})_{11}{\big)}}\\
&\leq k_1\cdot
\frac{4{\big|}(\chi^{\Delta_1} V \chi^{\Delta_1} )_{11}{\big|}^2}{(\chi^{{\Delta_1} \dagger}\chi^{\Delta_1})_{11}}.
\end{aligned}
\end{eqnarray*}
\end{widetext}
In general, in this situation the measurement rate per photon is not greater than the bound where there is only one drive coupling with mode $1$.

Secondly, consider the case where $|w_{\textsf{dr},1}- w_{\textsf{dr},2}|\gg |\Delta_2| $.
Similarly, under the RWA condition $$|w_{\textsf{dr},1}- w_{\textsf{dr},2}|\gg \max\{|\Delta_2|,~\|\tilde{H}\|,~k_i,\sqrt{k_i} \beta_i\}~(\text{for}~i=1,~2),$$
the signals containing the rapid oscillation $\sqrt{k_1}\beta_1 \exp\{i(\omega_{\textsf{dr},2}-\omega_{\textsf{dr},1})t\}$ from mode $1$ can be averaged out in the long-time limit.  From Eq.~(\ref{main3}), if $t$ is sufficiently large, the annihilation operator of the mode $l$ is described as
\begin{equation*}
\begin{aligned}
\hat{a}_l(t)=&\frac{\sqrt{k_2}}{i k_1}\beta_2\tilde{\chi}_{l2}^{\Delta_2}
+\frac{\sqrt{k_1}}{i k_1}\hat{B}^{\textsf{in}}_1\tilde{\chi}_{l1}^{\Delta_2}+\frac{\sqrt{k_2}}{i k_1}\hat{B}^{\textsf{in}}_2\tilde{\chi}_{l2}^{\Delta_2}\\
&+\frac{\sqrt{2}}{i k_1}{\big(}\sum^{N_Y}_{j=1}(\tilde{\chi}^{\Delta_2} Y)_{lj}C^{\textsf{in}\dagger}_j+\sum^{N_Z}_{j=1}(\tilde{\chi}^{\Delta_2} Z)_{lj}D^{\textsf{in}}_j{\big)}.
\end{aligned}
\end{equation*}

Following a similar  analysis to that in Appendix B, we have
\begin{equation}\label{signal2}
\mathcal{S}=2\epsilon^2\tau^2\frac{k_2}{k_1^2}\beta_2^2{\big|}(\chi^{\Delta_2} V \chi^{\Delta_2} )_{12}{\big|}^2,
\end{equation}
which only contains the  signals from  mode $2$.

The total number of photons is
\begin{equation}\label{ntot2}
\bar{n}_{\textsf{tot}}
=\frac{\beta_1^2}{k_1} (\chi^{\Delta_2^\prime \dagger}\chi^{\Delta_2^\prime})_{11}+\frac{k_2\beta_2^2 }{k_1^2} (\chi^{\Delta_2\dagger}\chi^{\Delta_2})_{22},
\end{equation}
where $\Delta_2^\prime=\Delta_2-\omega_{\textsf{dr},2}+\omega_{\textsf{dr},1}$. It contains the photons injected through mode $1$ and mode $2$.

Moreover, the noise power is described as
\begin{equation}\label{Noise-4}
\begin{aligned}
\mathcal {N}=&\frac{k_1 \tau}{2}{\Big (}1+\frac{4}{k_1}(\chi^{\Delta_2} YY^\dagger \chi^{\Delta_2\dagger})_{11}{\Big)}.
\end{aligned}
\end{equation}
From Eqs.~(\ref{signal2}), (\ref{ntot2}) and  (\ref{Noise-4}), the measurement rate per photon is described as
\begin{widetext}
\begin{eqnarray}\label{bound4}
\begin{aligned}
\bar{\Gamma}_{\textsf{meas}}&=k_1\cdot
\frac{4\frac{k_2}{k_1}|\frac{\beta_2}{\beta_1}|^2{\big|}(\chi^{\Delta_2} V \chi^{\Delta_2} )_{11}{\big|}^2}{{\big(}(\chi^{{\Delta_2^{\prime}} \dagger}\chi^{\Delta_2^{\prime}})_{11}+\frac{k_2}{k_1}| \frac{\beta_2}{\beta_1}|^2 (\chi^{{\Delta_2} \dagger}\chi^{\Delta_2})_{22}{\big)}
{\big(}1+\frac{4}{k_1}(\chi^{\Delta_2} Y Y^\dagger \chi^{{\Delta_2} \dagger})_{11}{\big)}}\\
&\leq k_1\cdot
\frac{4\frac{k_2}{k_1}|\frac{\beta_2}{\beta_1}|^2{\big|}(\chi^{\Delta_2} V \chi^{\Delta_2} )_{11}{\big|}^2}{{\big(}(\chi^{{\Delta_2^{\prime}} \dagger}\chi^{\Delta_2^{\prime}})_{11}+\frac{k_2}{k_1}| \frac{\beta_2}{\beta_1}|^2 (\chi^{{\Delta_2} \dagger}\chi^{\Delta_2})_{22}{\big)}
{\big(}1+2\Theta(\Xi)\cdot\Xi{\big)}},
\end{aligned}
\end{eqnarray}
\end{widetext}
where $\Xi(\Delta_2)=-(\chi^{\Delta_2}_{11}+\chi^{{\Delta_2}\ast}_{11})+|\chi_{11}^{\Delta_2}|^2+\frac{k_2}{k_1}|\chi_{12}^{\Delta_2}|^2$.

For two-mode sensors, if the coupling matrix is  $V=\frac{1}{2}\sigma_x$
as in Section IV, from Eq.~(\ref{bound4}) we have the best possible measurement rate per photon as
\begin{widetext}
\begin{eqnarray*}\label{bound5}
\bar{\Gamma}_{\textsf{2-opt}}= k_1\cdot
\frac{\frac{k_2}{k_1}|\frac{\beta_2}{\beta_1}|^2{\big|}{\chi_{12}^{\Delta_2}}^2+  \chi^{\Delta_2}_{11}\chi^{\Delta_2}_{22}{\big|}^2}{{\big(}|\chi_{11}^{\Delta^{\prime}_2}|^2+|\chi_{21}^{\Delta^{\prime}_2}|^2
+\frac{k_2}{k_1}|\frac{\beta_2}{\beta_1}|^2(|\chi_{12}^{\Delta_2}|^2+|\chi_{22}^{\Delta_2}|^2){\big)}
{\big(}1+2\Theta(\Xi)\cdot\Xi{\big)}}.
\end{eqnarray*}
\end{widetext}
Since we have assumed that $$|w_{\textsf{dr},1}-w_{\textsf{dr},2}|\gg 1,$$ when $|\chi_{12}^{\Delta_2}|$ becomes large, the elements of the transfer matrix $\chi^{\Delta^\prime_2}$ keep small. Thus it can be verified that if  conditions (\ref{cond1-a}) and (\ref{cond2-a}) hold with $\Delta$ being replaced by $\Delta_2$, then
$$\bar{\Gamma}_{\textsf{2-opt}}=\frac{1}{2}\cdot\frac{k_1}{k_2}\cdot k_1.$$
This means that the excitation from mode 2 dominates in the total signal power and the signal power gain from mode $2$ prevails in the competition with the noise enhancement induced by the unavoidable associated noise.

\end{document}